\documentclass[aps,preprint,tightenlines,superscriptaddress,showpacs]{revtex4}
\usepackage{amsmath}
\usepackage{epsfig}
\newcommand{\bra}{\langle}
\newcommand{\ket}{\rangle}

\newcommand{\bs}[1]{\ensuremath{\boldsymbol{#1}}}
\begin{document}
\title{Are there compact heavy four-quark bound states?}

\author{J. Vijande}
\affiliation{
Departamento de F\' \i sica Te\'orica e IFIC,
Universidad de Valencia - CSIC, E-46100 Burjassot, Valencia, Spain}
\author{E. Weissman}
\affiliation{The Racah Institute of Physics, The Hebrew University, 91904,
Jerusalem, Israel}
\author{A. Valcarce}
\affiliation{Departamento de F\'\i sica Fundamental, Universidad de Salamanca, E-37008 Salamanca, Spain}
\author{N. Barnea}
\affiliation{The Racah Institute of Physics, The Hebrew University, 91904,
Jerusalem, Israel}
\affiliation{Institute for Nuclear Theory, University of Washington,
Seattle, WA 98195, USA}

\date{\today}

\begin{abstract}
We present an exact method to study four-quark systems based on the
hyperspherical 
harmonics formalism. We apply it to several physical systems of interest 
containing two heavy and two light quarks using different
quark-quark potentials. Our conclusions mark the boundaries for
the possible existence of compact, non-molecular, four-quark bound states. While
$QQ\bar n \bar n$ states may be stable in nature, the stability
of $Q\bar Qn \bar n$ states would imply the existence of quark
correlations not taken into account by simple quark dynamical models.
\end{abstract}

\maketitle

The discoveries on several fronts~\cite{Pdg06}, of unusual 
charmonium states like $X(3872)$ and $Y(4260)$ and 
open-charm mesons with unexpected masses like
$D_{sJ}^*(2317)$ and $D^*_0(2308)$, 
have re-invigorated the study of the hadron spectra. 
Their anomalous nature has triggered
several interpretations, among them, the existence of 
four-quark states or meson-meson molecules. This challenging situation
resembles the long-standing problem of the light-scalar
mesons, where it has been suggested that some resonances
may not be ordinary $q \bar q$ states, though
there is little agreement on what they actually are~\cite{Ams04}.
In this case, four-quark states have been justified to coexist 
with $q\bar q$ states because they can couple to $J^{PC}=0^{++}$ without
orbital excitation~\cite{Jaf05}.

Any debate on the possible multiquark structure of meson
resonances should be based on our
capability to find an exact solution of the four-body 
problem~\cite{Ade82}. Theoretical predictions
often differ because of the approximation method used.
A powerful tool to solve a few-particle system is to
expand the trial wave function in terms of hyperspherical harmonics (HH)
basis functions. This method has been proven to be rather powerful to solve
the nuclear~\cite{pisa} 
four-body problem. In this work we use a generalization of the 
HH formalism to study four-quark systems
in an exact way. There are two basic difficulties for 
constructing HH functions of proper symmetry for a system of identical particles:
first, the simultaneous treatment of particles and antiparticles,
and second the additional color and flavor degrees of freedom.
The method will be tested by comparing with the existing results
based on different approximate solutions, thus establishing the validity 
of such approximations. Due to their actual interest and having in mind that systems with
unequal masses are more promising to be bound~\cite{Ade82}, we will center on the study of
$Q Q\bar n \bar n$ and $Q  \bar Q n\bar n$ states ($n$ stands
for a light quark and $Q$ for a heavy one). We will analyze the possible
existence of compact four-quark bound states using two standard quark-quark
interactions, a Bhaduri-like potential (BCN)~\cite{Bha81} and
a constituent quark model considering boson
exchanges (CQC) \cite{Vij05}. Both interactions
give a reasonable description of the meson and the baryon spectroscopy,
a thoughtful requirement considering that in the tetraquarks $qq$ and $q\bar q$
interactions will contribute.

Within the HH expansion, the four--quark wave function 
can be written as a sum of outer products 
of color, isospin, spin and configuration terms
\begin{equation}
    |\phi_{CISR}\ket= |{\rm Color}\ket |{\rm Isospin}\ket
               \left[|{\rm Spin}\ket \otimes| R \ket \right]^{J M} \, ,
\end{equation}
such that the four-quark state is a color singlet with well defined
parity, isospin and total angular momentum.
In the following we shall assume that particles $1$ and $2$ are the $Q$-quarks
and particles $3$ and $4$ are the $n$-quarks.
In the $QQ\bar n \bar n$ case
particles 1 and 2 are identical, and so are 3 and 4.
Consequently, the Pauli principle leads to the following 
conditions,
\begin{equation}\label{pauli}
\hat P_{12}|\phi_{CISR}\ket=\hat P_{34}|\phi_{CISR}\ket=-|\phi_{CISR}\ket \, ,
\end{equation}
$\hat P_{ij}$ being the permutation operator of particles $i$ and $j$.

Coupling the color states of two quarks (antiquarks) can yield two possible
representations, the symmetric $6$-dimensional, $6$ ($\bar 6$),
and the antisymmetric $3$-dimensional, $\bar 3$ ($3$).
Coupling the color states of the quark pair with that of the antiquark pair
must yield a color singlet. Thus, there are only two possible color states for a
$QQ\bar q \bar q$ system \cite{Jaf77},
\begin{equation}
  |{\rm Color}\ket = \{ | \bar 3_{12} 3_{34} \ket , | 6_{12} \bar 6_{34} \ket\}\,.
\end{equation}
These states have well defined symmetry under permutations, Eq. (\ref{pauli}).
The spin states with such symmetry can be obtained in the following way,
\begin{equation}
  |{\rm Spin}\ket = |((s_1,s_2)S_{12},(s_3,s_4)S_{34})S\ket
                  = | (S_{12} S_{34}) S \ket\;.
\end{equation}
The same holds for the isospin, $|{\rm Isospin}\ket=|(i_3,i_4)I_{34} \ket$,
which applies only to the $n$-quarks, thus $I=I_{34}$.

As mentioned above, we use the HH expansion to describe the spatial
part of the wave function. We choose for convenience the $H$-type 
Jacobi coordinates, 
\begin{eqnarray}
\bs{\eta}_1 & =&  \mu_{1,2}(\bs r_2 - \bs r_1) \, , \cr
\bs{\eta}_2 & =&  \mu_{12,34} 
                 \left(\frac{m_3\bs r_3+m_4\bs r_4}{m_{34}} 
                      -\frac{m_1\bs r_1+m_2\bs r_2}{m_{12}} \right) \, , \cr
\bs{\eta}_3 & =&  \mu_{3,4}(\bs r_4 - \bs r_3) \, ,
\end{eqnarray}
were $m_{ij}=m_i+m_j$, $\mu_{i,j}=\sqrt{m_i m_j/m_{ij}}$, and $m_{1234}=m_1+m_2+m_3+m_4$.
Using these vectors, it is easy
to obtain basis functions that have well defined symmetry under permutations
of the pairs $(12)$ and $(34)$. In the HH formalism the three Jacobi vectors 
are transformed into a single length variable,
$\rho=\sqrt{\eta_1^2+\eta_2^2+\eta_3^2}$, and $8$-angular variables, $\Omega$, that
represent the location on the $8$-dimensional sphere. 
The spatial basis states are given by 
\begin{equation}
\bra \rho \Omega |R\ket = U_{n}(\rho){\cal Y}_{ [K]} (\Omega) \, , 
\end{equation}
were ${\cal Y}_{ [K]} $ are the HH functions, and 
$[K]\equiv \{K K_{12} L M_L L_{12} \ell_3 \ell_2 \ell_1\}$.
The quantum number $K$ is the grand angular momentum, 
$L M_L$ are the usual orbital angular momentum
quantum numbers, and $\ell_i$ is the angular momentum associated
with the Jacobi vector $\bs{\eta}_i$. The quantum
numbers $K_{12}, L_{12}$ correspond to the intermediate coupling of
$\bs{\eta}_1$ and $\bs{\eta}_2$.
The Laguerre functions are used as the hyper--radial basis
functions $U_n(\rho)$.

The Pauli principle, Eq. (\ref{pauli}), leads to the following restrictions
on the allowed combinations of basis states:
\begin{itemize}
\item[{(i)}] $(-1)^{S_{12}+\ell_1}=+1$, $(-1)^{S_{34}+I+\ell_3}=-1$ 
for the $| 6_{12} \bar 6_{34} \ket$ color state,
\item[{(ii)}] $(-1)^{S_{12}+\ell_1}=-1$, $(-1)^{S_{34}+I+\ell_3}=+1$ for 
the $| \bar 3_{12} 3_{34} \ket$ state.
\end{itemize}
In the $Q \bar Q  n \bar n$ case the particle 2 is the antiparticle of the particle 1, 
and the particle 4 is the antiparticle of the particle 3.
Assuming that $C-$parity is a good symmetry of QCD we can regard quarks and
antiquarks as identical particles and impose the symmetry condition,
Eq. (\ref{pauli}), on the $Q \bar Q  n \bar n$ system as well.
Coupling the color states of a quark and an antiquark can yield two possible
representations: the singlet and the octet.
These representations should be combined in the following way \cite{Jaf77}
$\{|1_{12}1_{34}\ket, | 8_{12}, 8_{34}\ket \}$ to yield a total color singlet state.
However, these states have not definite symmetry under particle
permutations $(12)$ and $(34)$.
To construct symmetrized states for the $Q\bar Q$ pair we
consider the following combinations,
\begin{equation}
 | C_{12}^{\Gamma_{12}} \ket = \frac{1}{\sqrt 2} ( | C_{12} \ket + \Gamma_{12}
   | C_{21}\ket  \, ,
\end{equation}
were $C_{12}=\{1_{12},8_{12}\}$, and $\Gamma_{12}=+1$ for a symmetric
combination and $-1$ for an antisymmetric one. For light quarks the
color and isospin states should be combined together to form states with well
defined symmetry. For $I_z=0$, for instance, these states take the form, 
\begin{equation}
|( C_{34}\, I_{34})^{\Gamma_{34}}\ket  =
    +\frac{_1}{^2}\left[|C_{34}\ket\left(|u \bar u\ket\pm |d \bar d\ket\right)
    +\Gamma_{34}|C_{43}\ket\left(|\bar u u\ket \pm |\bar d d\ket\right)\right] ,
\end{equation}
were the plus sign stands for $I_{34}=0$ state and the minus sign for the
$I_{34}=1$ state. As before, $C_{34}$ stands for either the singlet or the
octet representations. The total color-isospin states, 
$| C_{12}^{\Gamma_{12}} ( C_{34}\, I_{34})^{\Gamma_{34}} \ket$ are not only
good symmetry states, but also good 
$C-$parity states with,
$C=\Gamma_{12}\Gamma_{34}$. Imposing the Pauli principle for the $Q\bar Q n
\bar n$ system we get the following restrictions:
$\Gamma_{12}(-1)^{S_{12}+\ell_1}=+1$, $\Gamma_{34}(-1)^{S_{34}+\ell_3}=+1$, on
the basis states.

Assuming non-relativistic quantum 
mechanics we solve the four-body Schr\"odinger 
equation using the basis states described above. The grand angular momentum
$K$ is the main quantum number in our expansion and the
calculation is truncated at some $K$ value.
As mentioned above, 
for our study we will use two standard quark potential models 
providing a reasonable description of the hadron spectra.
In the following we draw the basic properties of the interacting potentials. 

The BCN model was proposed in the early 80's by Bhaduri {\it et al.} in an
attempt to obtain a  
unified description of meson and baryon spectroscopy \cite{Bha81}. It was
later on applied to  
study the baryon spectra \cite{Sil85} and four-quark ($q q \bar q \bar q$)
systems \cite{Sil93}.  
The model retains the 
most important terms of the one-gluon exchange interaction proposed by de
R\'ujula {\it et al.} \cite{Ruj75},  
namely coulomb and spin-spin terms, and a linear confining potential, having
the form 
\begin{equation}
{V(\vec r_{ij})=-\frac{3}{16}(\vec\lambda^c_i \cdot \vec\lambda^c_j)}
\times \left(\frac{r_{ij}}{a^2}-\frac{\kappa}{r_{ij}}-D+
\frac{\kappa}{m_im_j}\frac{e^{-r_{ij}/r_0}}{r_{ij}r_0^2}(\vec\sigma_i \cdot \vec\sigma_j)\right)\,,
\end{equation}
where $\vec \sigma_i$ are the Pauli matrices and $\vec \lambda^c_i$ are the $SU(3)$ color matrices. 
The parameters
$\kappa=102.67$ MeV fm, $D$=913.5 MeV, $a=$0.0326 MeV$^{-1/2}$ fm$^{1/2}$, $r_0=2.2$ fm, 
$m_{u,d}=$337 MeV, and $m_c=1870$ MeV are taken from Ref. \cite{Sil93}.

The CQC  model was proposed in the early 90's in an attempt to
obtain a simultaneous description of the nucleon-nucleon
interaction and the baryon spectra \cite{Rep05}.
It was later on generalized to all flavor sectors giving a reasonable description
of the meson \cite{Vij05} and baryon spectra \cite{Vij04b}. The possible existence of four-quark states
within this model has also been addressed \cite{Vij05b,Bar06}. 

The model is based on the assumption that the light-quark 
constituent mass appears because of the spontaneous breaking of the original $SU(3)_{L}\otimes SU(3)_{R}$ 
chiral symmetry at some momentum scale. In this domain of momenta, quarks interact through 
Goldstone boson exchange potentials. 
QCD perturbative effects are taken into account through the one-gluon-exchange (OGE) potential 
as the one used in the BCN model. Finally, it incorporates 
confinement as dictated by unquenched lattice calculations
predicting, for heavy quarks, a screening effect on the 
linearly dependent interquark
potential when increasing the interquark distance \cite{Bal01}.

The model parameters have been taken from Ref. \cite{Vij05} with 
the exception of the OGE regularization parameter. This 
parameter, taking into account the size of the system, 
was fitted for four--quark states in the description of the light scalar sector \cite{Vij05b}, 
being $\hat r_0=0.18$ fm for mesons and $\hat r_0=0.38$ fm for four-quark systems. 
Let us also notice that the CQC model contains an 
interaction generating flavor mixing
between $n\bar n$ and $s \bar s$ components. It allows to
exactly reproduce the masses of the $\eta$ and $\eta'$ mesons.
In the four--quark case this contribution would mix isospin zero
$Q\bar Q n\bar n$ and $Q\bar Q s \bar s$ components. Such
contributions were explicitly evaluated in the variational
approach of Ref. \cite{Vij05b} for the light isocalar tetraquarks,
giving a negligible effect. In order 
to make a proper comparison between thresholds and four--quark states 
we have recalculated the meson spectra of Ref. \cite{Vij05} with 
the same $\hat r_0$ value and interaction used in the four-quark 
calculation, neglecting therefore the flavor--mixing terms. 
Explicit expressions of the interacting potentials and a more detailed 
discussion of the model can be found in Ref. \cite{Vij05}.
%
%%%%%%%%%%%%%%%%%%%%%%%%%%%%%%%%%%%%%%%%%%%%%%%%%%%%%%%%%%%%%%%%%%%%%
\begin{table}[t]
\caption{Energy (MeV) of $L=0$ $cc\bar n\bar n$ states.}
\label{t0}
\begin{tabular}{|c|ccc|}
\hline
(S,I)   & Ref. \cite{Vij04}   & HH($ \ell_i=0$)& HH\\
\hline
(0,1)   & 4155          & 4154  & 3911   \\
(1,0)   & 3927          & 3926  & 3860   \\
(1,1)   & 4176          & 4175  & 3975   \\
(2,1)   & 4195          & 4193  & 4031   \\
\hline
\end{tabular}
\end{table}
%%%%%%%%%%%%%%%%%%%%%%%%%%%%%%%%%%%%%%%%%%%%%%%%%%%%%%%%%%%%%%%%%%%%%

\begin{table}[b]
\caption{Energy (MeV) of $L=0$ $c \bar c n \bar n$ states.}
\label{t1}
\begin{tabular}{|c|ccc|}
\hline
$J^P$&Ref. \cite{Sil93}& HH$(K=8)$& HH$(K_{\rm max})$ \\ 
\hline
$0^+$&3409& 3380& 3249 $(26)$\\
$1^+$&3468& 3436& 3319 $(22)$\\
\hline
\end{tabular}
\end{table}
%%%%%%%%%%%%%%%%%%%%%%%%%%%%%%%%%%%%%%%%%%%%%%%%%%%%%%%%%%%%%%555

Let us first analyze the numerical capability of the designed method to 
capture the properties of the four-quark systems.
In Table \ref{t0} we present the results for different $L=0$ spin-isospin
$cc \bar n\bar n$ states calculated with the CQC model. 
We quote in the first column the
results obtained within a variational calculation
using Gaussian trial wave functions only with quadratic terms in the
Jacobi coordinates \cite{Vij04}. This approximation would correspond
in our formalism to set $\ell_i=0$ for the three Jacobi vectors. 
These results are given in the second column, reproducing 
exactly the variational results.
The validity of this approximation can be judged
by looking at the last column where we give the 
exact HH results, truncated at $K=24$. In some cases the difference between
the $\ell_i=0$ approximation and the true ground state can be as
large as $200$ MeV. We have also reproduced the calculation of the
$(S,I)=(1,0)$ $L=0$ $c c\bar n\bar n$ state of Refs. \cite{Jan04,Sil93}
using the BCN model. For $K=24$ we have obtained an
energy of 3899.7 MeV as compared to 3904.7 MeV of Ref. \cite{Jan04}
and 3931.0 MeV of Ref. \cite{Sil93}. Ref. \cite{Jan04} designed
a powerful method, similar to the stochastic variational approach \cite{Varga},
to study this particular system. Although their results are not fully
converged, the close agreement gives confidence on both
calculations. The results of Ref. \cite{Sil93} were obtained using
diagonalization in harmonic oscillator (HO) basis up to 
to $N=8$. The quality of this last procedure can be judged by looking
at Table \ref{t1} where we compare, for different $c \bar c n \bar n$
states, the results of \cite{Sil93}
to the HH results with $K \leq 8$ and to the converged HH results
obtained with $K$ restricted by our computational capability,
i.e. $K=22$ or $K=26$.  
As can be seen the results with the basis truncated at $K=8$ are similar
to the HO results, but rather far from the converged ones.

%%%%%%%%%%%%%%%%%%%%%%%%%%%%%%%%%%%%%%%%%%%%%%%%%%%%%%%%%%%%%%%%5
\begin{table}[t]
\caption{Energy (MeV) and probability of the different 
color components as a function of $K$ for the $c \bar c n\bar n$
$J^{PC}=1^{++}$ state both for CQC and BCN models. 
The last rows indicate the lowest theoretical two-meson thresholds. The
notation $\vert_S$ ($\vert_P$) stands for relative $S$-wave ($P$-wave).
}
\label{t2}
\begin{tabular}{|c||ccc|ccc|}
\hline
&\multicolumn{3}{|c|}{CQC}&\multicolumn{3}{|c|}{BCN}\\
\hline
$K$     & E   & $P_{11}$&$P_{88}$& E    & $P_{11}$& $P_{88}$    \\
\hline
0       & 4141&  1.0000 & 0.0000& 4196  &1.0000 & 0.0000 \\
2       & 3985&  0.9822 & 0.0178& 4053  &0.9462 & 0.0538 \\
4       & 3911&  0.9789 & 0.0211& 3994  &0.9233 & 0.0767 \\
6       & 3870&  0.9834 & 0.0166& 3963  &0.9236 & 0.0764 \\
8       & 3845&  0.9871 & 0.0129& 3944  &0.9303 & 0.0697 \\
10      & 3827&  0.9905 & 0.0095& 3932  &0.9426 & 0.0574 \\
12      & 3814&  0.9926 & 0.0074& 3920  &0.9927 & 0.0073 \\
14      & 3805&  0.9943 & 0.0057& 3887  &0.9990 & 0.0010 \\
16      & 3797&  0.9954 & 0.0046& 3861  &0.9994 & 0.0006 \\
18      & 3791&  0.9962 & 0.0038& 3840  &0.9995 & 0.0005 \\
20      & 3786&  0.9968 & 0.0032& 3822  &0.9996 & 0.0004 \\
22      & $-$ &  $-$    &  $-$  & 3808  &0.9997 & 0.0003 \\
\hline
\hline
$J/\psi\,\omega\vert_S$ & 3745  & 1 & 0 & 3874 & 1 & 0  \\
$\chi_{cJ}\,\eta \vert_P$ & 4281  & 1 & 0 & 3655 & 1 & 0  \\
\hline
\end{tabular}
\end{table}
%%%%%%%%%%%%%%%%%%%%%%%%%%%%%%%%%%%%%%%%%%%%%%%%%%%%%%%%%%%%%%

In spite of the shortcomings of the methods used to study
four-quark systems, in the past, many four-quark bound states 
have been suggested. 
To analyze their stability against dissociation,
the parity and the total angular momentum 
must be preserved. Additionally, $C-$parity
is a good quantum number for $c\bar c n\bar n$
and the Pauli principle must be fulfilled in the $cc\bar n\bar n$ case.
The thresholds can be evaluated by adding the meson masses
of the dissociation channel. 
A four-quark state will be stable under strong 
interaction, and therefore narrow, if its mass 
lies below all allowed two-meson thresholds. 
Sometimes, the results of four-quark calculations have been directly compared to
the experimental thresholds. In this case 
one could misidentify scattering wave functions as bound states.
When they are referred to the threshold within the same model,
we will see how the theoretical predictions do not imply an
abundance of multiquark states in the data.

Let us make a brief comment on the convergence of the HH expansion.
In some cases the convergence is slow, as can be seen by comparing 
Tables \ref{t2} and \ref{t3}. This is a consequence of the HH
formalism being better suited to treat with bound states, 
and most of the four-quark states one deals with are above the corresponding 
two-meson threshold, as can be seen in
Table \ref{t2}. Due to this slowness, our calculation 
cannot definitively exclude large molecular objects (sizes of 
the order of 1.5--2 fm) with binding energies smaller than 1--2 MeV 
induced by long-range interactions like, for instance, the one-pion exchange tail \cite{Tor04}.

%%%%%%%%%%%%%%%%%%%%%%%%%%%%%%%%%%%%%%%%%%%%%%%%%%%%%%%%%%%%%%%%%%
\begin{table}[t]
\caption{Same as Table \protect\ref{t2}
for the $cc\bar n \bar n$ $J^{P}=1^{+}$ state.}
\label{t3}
\begin{tabular}{|c||ccc|ccc|}
\hline
&\multicolumn{3}{|c|}{CQC}&\multicolumn{3}{|c|}{BCN}\\
\hline
$K$     & E   & $P_{11}$&$P_{88}$ & E   & $P_{11}$&$P_{88}$ \\
\hline
0       & 4109 & 0.3351      & 0.6649 &
4100    & 0.3446        & 0.6554 \\
2       & 3990 & 0.3483      & 0.6517 &
3999    & 0.3744        & 0.6256 \\
4       & 3931 & 0.3577      & 0.6423 &
3954    & 0.3981        & 0.6019 \\
6       & 3903 & 0.3641      & 0.6359 &
3933    & 0.4170        & 0.5830 \\
8       & 3887 & 0.3681      & 0.6319 &
3921    & 0.4302        & 0.5698 \\
10      & 3878 & 0.3705      & 0.6295 &
3914    & 0.4403        & 0.5597 \\
12      & 3872 & 0.3720      & 0.6280 &
3910    & 0.4478        & 0.5522 \\
14      & 3868 & 0.3730      & 0.6270 &
3907    & 0.4536        & 0.5464 \\
16      & 3866 & 0.3737      & 0.6263 &
3904    & 0.4581        & 0.5419 \\
18      & 3864 & 0.3741      & 0.6259 &
3903    & 0.4618        & 0.5382 \\
20      & 3862 &   $-$       &  $-$   &
3901    & 0.4647        & 0.5353 \\
22      & 3861 &   $-$       &  $-$   &
3900    & $-$ & $-$ \\
24      & 3861 &  $-$        &  $-$   &
3900    &    $-$        &  $-$  \\
\hline
$D\,D^*\vert_S$& 3937       &  1 & 0 &
 3906       &  1 & 0 \\
\hline
\end{tabular}
\end{table}
%%%%%%%%%%%%%%%%%%%%%%%%%%%%%%%%%%%%%%%%%%%%%%%%%%%%%%%%%%%%%%%%%%%%%%%

\begin{figure}
\epsfig{file=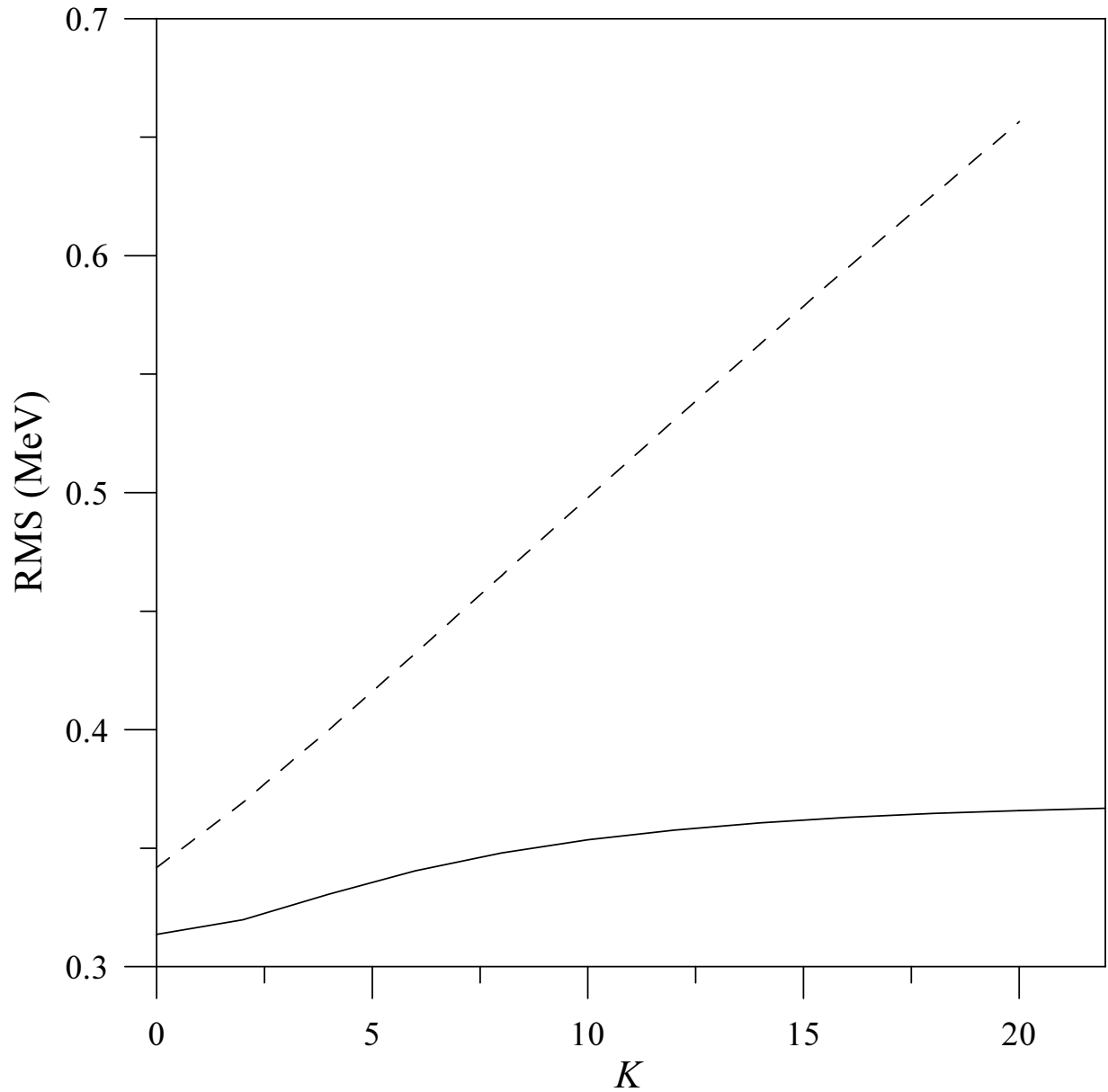, width=\linewidth}
\caption{Evolution with $K$ of the radius (RMS) of the $c\bar c n\bar n$
  $J^{PC}=1^{++}$ 
state (dashed line) and the $cc\bar n \bar n$ $J^P=1^+$ 
state (solid line) for the CQC model.
\label{f1}}
\end{figure}

Once the method has been designed, tested, and 
its accuracy established, we concentrate on a hot subject:
the $c\bar c n\bar n$ system as a potential structure 
for the $X(3872)$. To make the physics clear we will compare 
with the $cc \bar n \bar n$ system. In particular, we focus 
on the $J^{PC}=1^{++}$ $c\bar c n\bar n$ and 
$J^{P}=1^{+}$ $cc\bar n \bar n$ quantum numbers to illustrate
their similitude and differences. 
A complete study of all the quantum numbers 
will be reported elsewhere. 
The results are shown in Tables \ref{t2} and \ref{t3}
up to the maximum value of $K$ within our computational
capabilities. Since we are using a complete set of 
HH basis, all possible configurations are considered
in both cases. For the $c\bar c n\bar n$ system, independently
of the quark-quark interaction, the system evolves to a 
well separated two-meson state. This is clearly seen 
in the energy, approaching the corresponding two free-meson threshold, 
but also in the probabilities of the 
different color components of the wave function 
and in the radius. We denote by $P_{11}$ ($P_{88}$)
the probability of a singlet-singlet (octet-octet) color 
component in the $(c \bar c)(n \bar n)$ (or $(c \bar n)(c \bar n)$) coupling. 
We observe how the system evolves to 
two singlet color mesons, whose separation increases with $K$,
dashed line in Fig. \ref{f1}. 
Comparing the theoretical predictions with the experimental threshold, 
$M_{J/\psi\,\omega\vert_S} = 3879.57\pm0.13$ MeV,  
one could be tempted to claim for the existence of a bound state. However, the experimental
threshold is not reproduced by the effective Hamiltonians. In fact, in the BCN
model the sum of the masses of the two mesons $J/\psi\,\omega$ is even larger than 
that of $\chi_{cJ}\,\eta $, leading to a completely different threshold for the
$1^{++}$ system. Thus, in any manner one can claim for the existence
of a bound state.  Similar conclusions are drawn for 
all quantum numbers of this system.

A completely different behavior is observed in Table \ref{t3}.
Here, the energy is quickly stabilized below
the theoretical threshold. Besides, the radius
is also stable, solid line in Fig. \ref{f1}, and it is smaller than the
sum of the radius of the two-meson threshold. We obtain $r_{4q}=0.37$ fm
compared to $r_{M_1}+r_{M_2}= 0.44$ fm. We also notice a different
solution for the  probability of the color components.
However, one should not directly conclude
the presence of octet-octet components,
because the octet-octet color component in the 
$(c_1\bar n_3)(c_2\bar n_4)$ basis can be re-expressed
as a singlet-singlet color component in the
$(c_1\bar n_4)(c_2\bar n_3)$ coupling, being the
same physical system due to the identity of the two
quarks and the two antiquarks. Although in the BCN model
the system is slightly bound, the structure of the
bound state is manifest for low values of $K$, leading 
one to conclude that the state could hardly
be destroyed by small non-considered effects as
could be, for example, relativity. The actual interest
and the capability of some experiments~\cite{Sel05}
to detect double charm states makes this prediction
a primary objective to help in the understanding of QCD
dynamics.

%%%%%%%%%%%%%%%%%%%%%%%%%%%%%%%%%%%%%%%%%%%%%%%%%%%%%%%%%%%%%%%%%%%%5�
%%%%%%%%%%%%%%%%%%%%%%%%%%%%%%%%%%%%%%%%%%%%%%%%%%%%%%%%%%%%%%%%%%%%%

It is thus important to realize that a bound
state should be pursued not only by looking at the energy, but also
with a careful analysis of the radius and color probabilities. Besides,
one should compare results within the same calculating framework,
unless other effects, as discussed below, are considered.
This detailed analysis allows us to distinguish between 
compact states and meson-meson molecules \cite{Jaf05}
and it does consider the contribution of all meson-meson channels
to a particular set $J^{PC}$ of quantum numbers \cite{Tor04}.
Inherent to our discussion is a much richer decay spectrum
of compact states due to the presence of octet-octet color
components in their wave function.

Let us notice that
there is an important difference between the two physical systems studied.
While for the $c\bar c n\bar n$ there are two allowed physical {\it decay channels},
$(c\bar c)(n\bar n)$ and $(c\bar n)(\bar c n)$, for the $cc\bar n\bar n$ 
only one physical system contains the possible final states, $(c \bar n)(c\bar n)$. 
This has important consequences if both systems (two- and four-quark
states) are described within the same two-body Hamiltonian,
the $c \bar c n \bar n$ will hardly present bound states, because the system
will reorder itself to become the lightest two-meson state, either
$(c\bar c)(n\bar n)$ or $(c\bar n)(\bar c n)$. In other words,
if the attraction is provided by the interaction between
particles $i$ and $j$, it does also contribute to the asymptotic
two-meson state. This does not happen
for the $c c\bar n\bar n$ if the interaction between, for example,
the two quarks is strongly attractive. In this case there is no asymptotic two-meson
state including such attraction, and therefore the system will bind. 

Therefore, our conclusions can be made more general. If we have an $N$-quark
system described by two-body interactions in such a way that there exists
a subset of quarks that cannot make up a physical subsystem, then one may expect
the existence of $N$-quark bound states by means of central two-body potentials.
If this is not true one will hardly find $N-$quark bound states \cite{Lip75}. 
For the particular case of the tetraquarks,
this conclusion is exact if the confinement is described
by the first $SU(3)$ Casimir operator, because when the system is split
into two-mesons the confining contribution from the two isolated mesons
is the same as in the four-quark system.
The contribution of three-body color forces \cite{Dmi05}
would interfere in the simple
comparison of the asymptotic and the compact states.
Another possibility in the same line would be a modification
of the Hilbert space. If for some reason particular components of the
four-quark system (diquarks) would be favored against others, the
system could be compact \cite{Mai04}. Lattice QCD calculations \cite{Ale07}
confirm the phenomenological expectation that QCD dynamics favors the
formation of good diquarks \cite{Jaf05}, i.e., in the scalar positive
parity channel. However, they are large objects whose relevance to hadron
structure is still under study. All these alternatives will allow
to manage the four-quark system without affecting the threshold and
thus they may allow to generate any solution.

Let us finally note that in Ref. \cite{Vij07b} the stability of the $QQ\bar n\bar n$ and
$Q\bar Q n \bar n$ systems has been analyzed in a simple string model considering
only a multiquark confining interaction given by the minimum
of a flip-flop or a butterfly potential. The ground state of
systems made of two quarks and two antiquarks of equal masses
was found to be below the dissociation threshold. While for
the flavor exotic $QQ\bar n\bar n$ the binding increases when
increasing the mass ratio $m_Q/m_q$, for the cryptoexotic
$Q\bar Q n\bar n$ the effect of symmetry breaking is the opposite,
the system being unbound whenever $m_Q/m_q >1$. Although more 
realistic calculations are needed before establishing a
definitive conclusion, the findings of Ref. \cite{Vij07b} 
strengthened our results.

\vspace{2cm}
We thank Dr. L. Sommovigo for a careful reading of the manuscript.
This work has been partially funded by Ministerio de Ciencia y 
Tecnolog\'{\i}a under Contract No. FPA2004-05616, and by 
Junta de Castilla y Le\'{o}n under Contract No. SA016A07.

%%%%%%%%%%%%%%%%%%%%%%%%%%%%%%%%%%%%%%%%%%%%%%%%%%%%%%%%%%%%%%%%%%%%%%%%%%%%%
\end{document}